\documentclass[aps,prd,twocolumn,nofootinbib,
preprintnumbers,showkeys,
superscriptaddress]{revtex4-2}

\usepackage{graphicx}
\usepackage{dcolumn}
\usepackage{bm}
\usepackage{amsmath}
\usepackage{wasysym}
\usepackage{float}
\usepackage{multirow}
\usepackage{xcolor}
\usepackage{scrextend}
\usepackage[colorlinks=true, 
pdfstartview=FitV, 
bookmarks=true, 
bookmarksnumbered=true, 
breaklinks]{hyperref}
\usepackage{color}
\definecolor{blue}{rgb}{0.0, 0.0, 1.0}
\definecolor{red}{rgb}{1.0, 0.0, 0.0}
\definecolor{royalblue}{rgb}{0.0, 0.14, 0.4}
\hypersetup{linkcolor=royalblue, 
	citecolor=blue, 
	urlcolor=royalblue}

\def\orcid#1{\kern .08em\href{https://orcid.org/#1}{\includegraphics[keepaspectratio,width=0.7em]{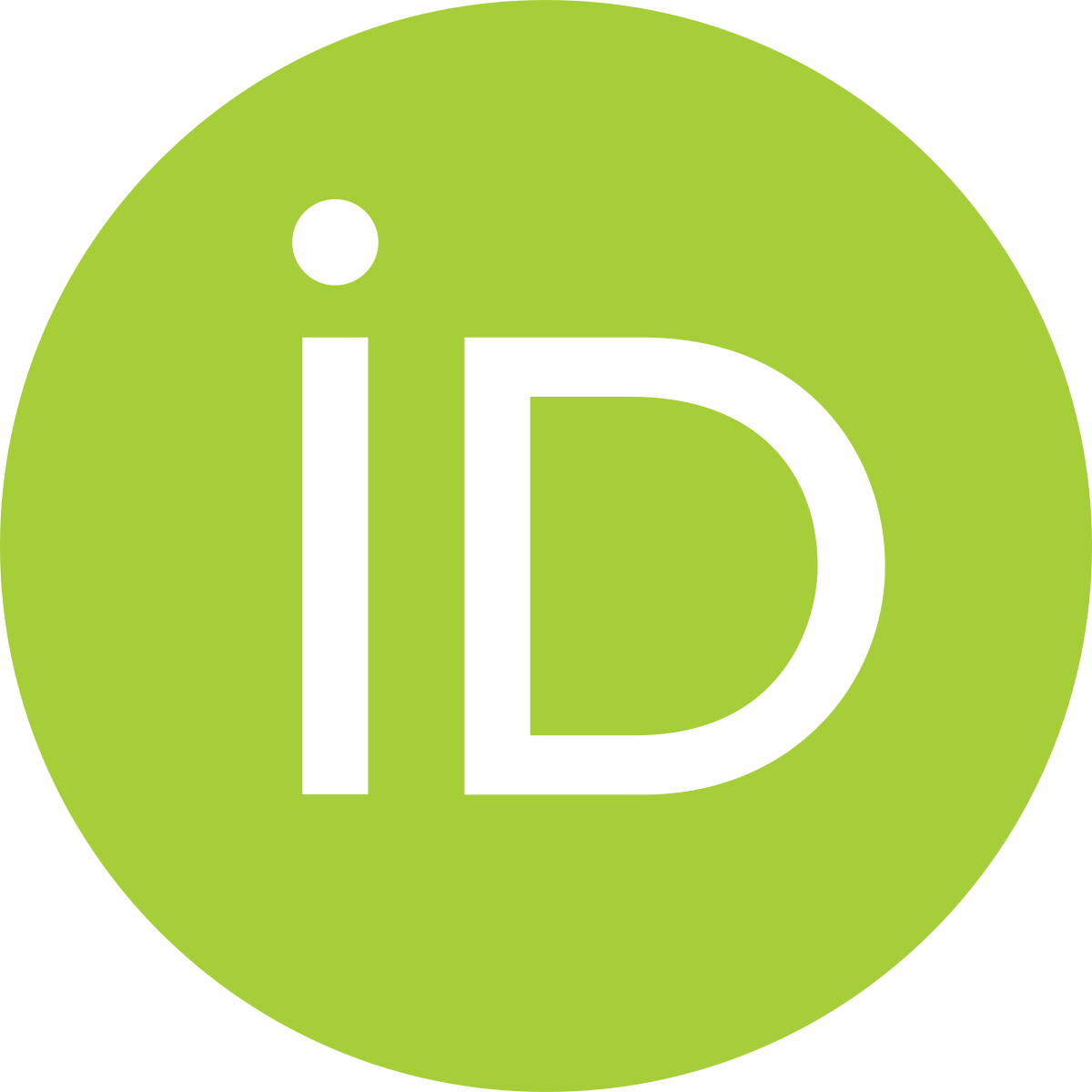}}}

\begin{document}
\title{Model independent analysis of coupled-channel scattering: \\
	a deep learning approach}

\author{Denny Lane B. Sombillo\orcid{0000-0001-9357-7236}}
\email[]{sombillo@rcnp.osaka-u.ac.jp}
\email[]{dbsombillo@up.edu.ph}
\affiliation{National Institute of Physics, University of the Philippines Diliman, Quezon City 1101, Philippines}
\affiliation{Research Center for Nuclear Physics (RCNP), Osaka University, Ibaraki, Osaka 567-0047, Japan}
%\orcid{0000-0001-9357-7236}
\author{Yoichi Ikeda\orcid{0000-0002-2235-1464}}
\affiliation{Department of Physics, Kyushu University, Fukuoka 819-0395, Japan}
%\orcid{0000-0002-2235-1464}
\author{Toru Sato\orcid{0000-0001-5216-5657}}
\affiliation{Research Center for Nuclear Physics (RCNP), Osaka University, Ibaraki, Osaka 567-0047, Japan}
%\orcid{0000-0001-5216-5657}
\author{Atsushi Hosaka\orcid{0000-0003-3623-6667}}
\affiliation{Research Center for Nuclear Physics (RCNP), Osaka University, Ibaraki, Osaka 567-0047, Japan}
\affiliation{Advanced Science Research Center, Japan Atomic Energy Agency, Tokai, Ibaraki 319-1195, Japan}
%\orcid{0000-0003-3623-6667}

\date{\today}

\begin{abstract}
	We develop a robust method to extract the pole configuration of a given partial-wave amplitude. In our approach, a deep neural network is constructed where the statistical errors of the experimental data are taken into account. The teaching dataset is constructed using a generic S-matrix parametrization, ensuring that all the poles produced are independent of each other. 
	The inclusion of statistical error results into a noisy classification dataset which we should solve using the curriculum method. 
	As an application, we use the elastic $\pi N$ amplitude in the $I(J^P)=1/2(1/2^{-})$ sector where
	$10^6$ amplitudes are produced by combining points in each error bar of the experimental data. 
	We fed the amplitudes to the trained deep neural network and find that the enhancements in the  $\pi N$ amplitude are caused by one pole in each nearby unphysical sheet and at most two poles in the distant sheet. 
	Finally, we show that the extracted pole configurations are independent of the way points in each error bar are drawn and combined, demonstrating the statistical robustness of our method.
\end{abstract}		
\maketitle

\section{Introduction}
\hspace{\parindent}
Identifying the nature of observed enhancements in hadron-hadron scatterings is one of the central goals of hadron spectroscopy~\cite{Olsen2018,Guo2018}. The primary concern is to associate the enhancements with nearby S-matrix poles~\cite{Taylor,Newton,EdenAnalytic,Pc4312,DongBaruGuoHanhartNefediev2020}. Ideally, we make a complete measurement of observables, such as elastic and reaction cross-sections with spin dependence if possible. For low energy processes, the measurement is followed by partial wave analysis of the experimental data to obtain the amplitudes. The poles, Riemann sheets, and other resonance parameters are extracted from the amplitudes. To achieve this program, conventionally, one often uses a simple parametrization of the amplitude such as Breit-Wigner or Flatte parametrizations, or some more rigorous formulation like the K-matrix formalism~\cite{BurkertLee2004}. Alternatively, one can use a dynamical model using suitable hadron degrees of freedom and their interactions. In these methods, some parameters are determined to fit the experimental data. In general, however, the applicability of these methods is somewhat limited by the assumptions of the model.

In this study, we propose an alternative method to the above parameter-fitting approach using deep learning~\cite{MLandPS,Sombillo2020,SombilloPRL2021}. 
Eventually, we would aim at the program that determines detailed properties of poles of the amplitude, including their positions, residues, among others. However, this is extremely difficult at this moment, and therefore, we limit our program to identify the pole configuration, that is, the number of poles in each Riemann sheet associated with the structures in the amplitude. The information we obtain is limited but valuable enough to draw some physical insights (see Refs.~\cite{Morgan1992,MorganPennington1993,Hyodo2002,Magas2005}). 

We design and develop a deep neural network (DNN) to detect the pole configuration using only the partial-wave amplitude. The idea is to produce a general set of simulated amplitudes with known pole configuration to build the training dataset. In doing so, the general properties of the S-matrix (analyticity, unitarity, and hermiticity) are satisfied. Furthermore, the poles that we generate must be independent of each other to ensure that the detected poles by the DNN are free from any bias. The next step is to construct a DNN program. In the optimization of the DNN, we use the training dataset to teach the DNN to recognize physically realizable amplitudes from the experimental data and extract the pole configuration. As anticipated, our method is a classification program that gives a general description of the data. Therefore, to extract the physical properties of the enhancement structures in the amplitudes, a dynamical model approach must be used while maintaining the result obtained by our program. We emphasize that once the DNN model we have constructed, the optimized parameters in DNN can be reused to analyze different processes. This treatment is possible since our formulation does not rely on a specific functional form of amplitudes.

\begin{figure}
	\includegraphics[width=0.95\columnwidth]{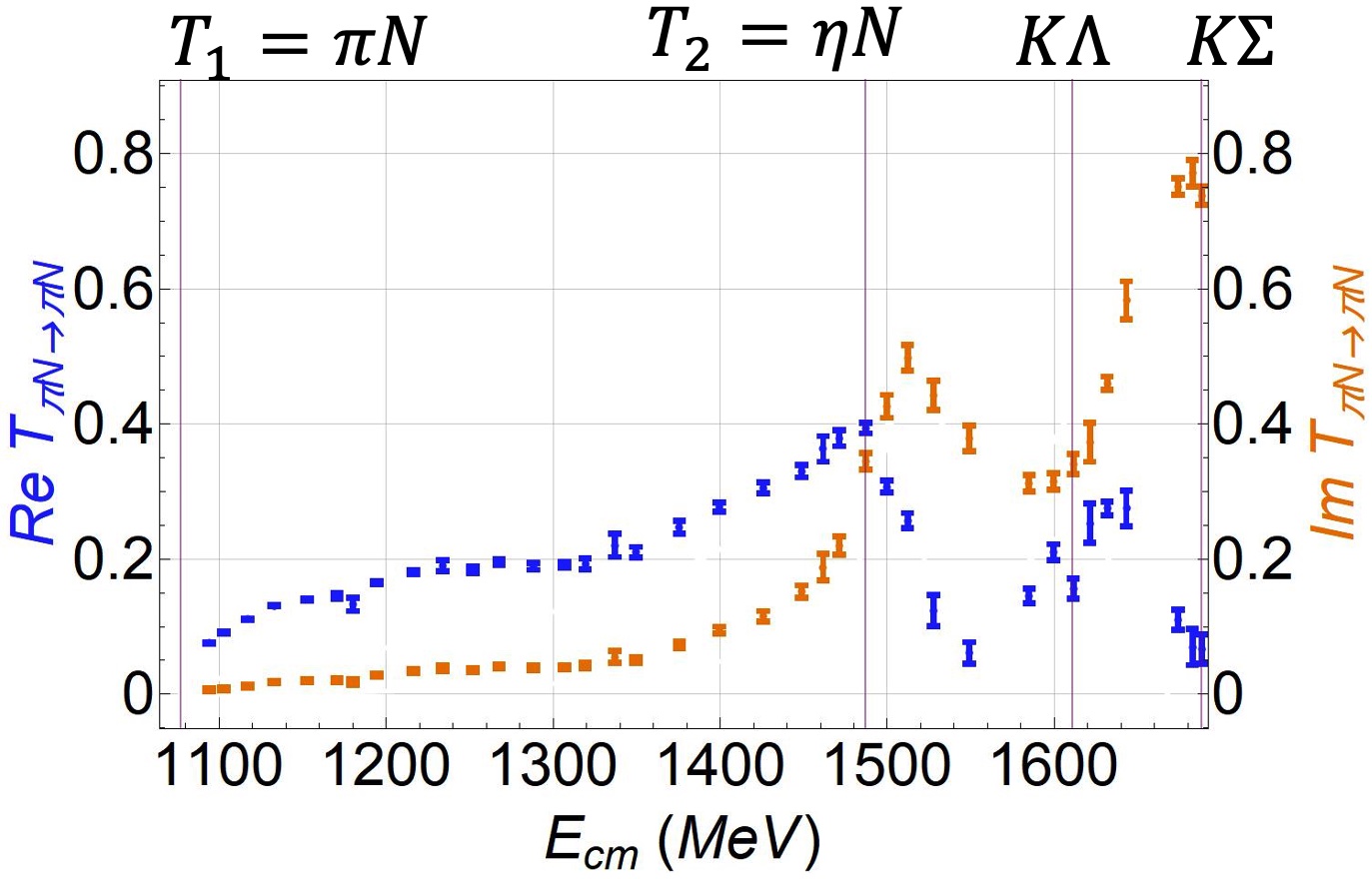}
	\caption{The elastic $\pi N$ amplitude of the GW-SAID in~\cite{GW_SAID}. 
		The two-hadron thresholds are shown as vertical thin lines. 
		Only the $\pi N$ and $\eta N$ channels are considered in the present study. 
		The $K\Lambda$ and $K\Sigma$ thresholds are shown for reference purposes only.}
	\label{fig:said_data}
\end{figure}

We consider the two-hadron scatterings with two coupled channels.
For a more specific demonstration, we take the elastic $\pi N$ partial-wave amplitude, collected and analyzed by the GW-SAID in Refs.~\cite{SAIDpiN,SAIDpiNold}, as our experimental data. We choose this amplitude because of the interesting role of threshold effects to the noticeable structures seen just above the $\eta N$ and below the $K\Sigma$ thresholds as shown in Fig.~\ref{fig:said_data}. We use the coupled $\pi N$ and $\eta N$ channels treatment for the present study. 
In addition to the deep learning approach, we also introduce an alternative way to utilize the error bars to interpret experimental data. Instead of generating several parameters of a model that fits within the experimental result, we do the reverse treatment. We produce several amplitudes directly from the experimental results by combining points in the error bars then feed them to the trained DNN. Since there are infinitely different ways to combine points in each error bar, we expect to get different pole configurations. We can interpret the DNN inference with the highest count as the best pole configuration associated with the experimental data. Thus, we are extracting the information directly from the experimental data and not just constraining the model. 

This paper serves as the companion paper of Ref.~\cite{SombilloPRL2021}. The structure is as follows. First, we discuss in section~\ref{sec:smatrix} the general properties of the S-matrix used to generate the training dataset. Then, we proceed with the construction and optimization of our DNN in section~\ref{sec:dnnmodel}. The treatment of experimental data for the DNN inference is done in section~\ref{sec:apply}. Finally, we give our conclusion in section~\ref{sec:conc}.

\section{General Properties of S-matrix}
\label{sec:smatrix}
\hspace{\parindent}
In this section, we review the general properties of the S-matrix that we will use in the generation of the training dataset. For the present purpose, it will suffice to consider the two-hadron scatterings with two channels. Here, we take into account the $\pi N$ as channel 1 and $\eta N$ as channel 2. The $K\Lambda$ channel can be ignored due to its weak coupling to the $\pi N$ channel~\cite{DCCKNLS2013}. We can also ignore the $K\Sigma$ channel by restricting the energy region of analysis up to the $K\Sigma$ threshold energy. 
The proximity of the peak structures to $\eta N$ threshold allows us to utilize the non-relativistic relation
\begin{equation}
	E = \dfrac{p_i^2}{2\mu_i}+T_i.
	\label{eq:nonrelE}
\end{equation}
where $E$ is the center-of-mass scattering energy, $p_i$ is the relative momentum of the $i$the channel ($i\in\{1,2\}$) with two-particle reduced mass $\mu_i$ and threshold energy $T_i$. Here, the smooth variation of the lower-channel is approximated using $p_1=\sqrt{2\mu_1(E-T_1)}$.

\begin{figure}
	\includegraphics[width=0.75\columnwidth]{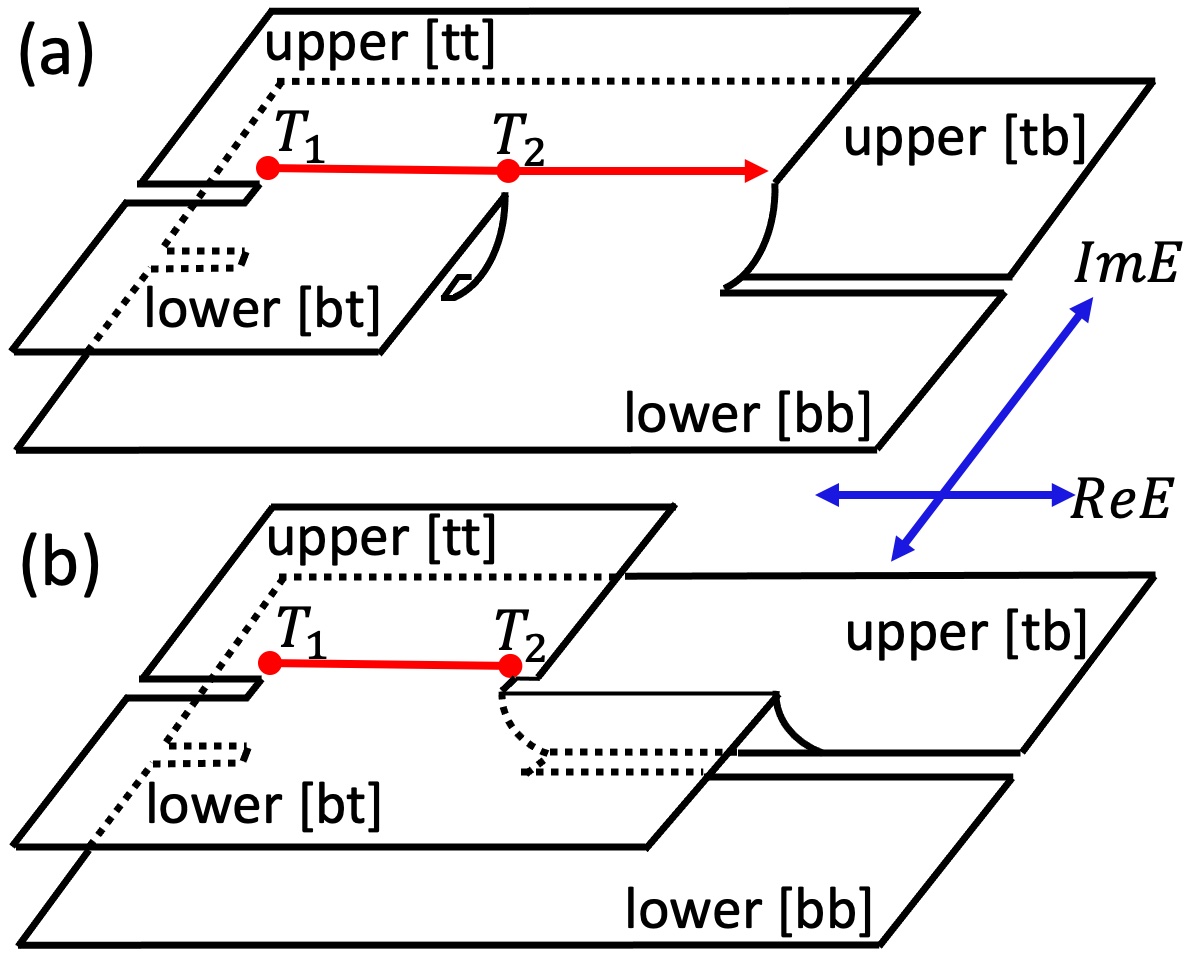}
	\caption{Relevant regions of two-channel Riemann sheets. The red ray, with two dots, represents the scattering region lying on the upper $[tt]$. The curve surface in (a) shows the $[tt]$-$[bb]$ connection, while the curve in (b) shows the $[bt]$-$[tb]$ connection.}
	\label{fig:riemann}
\end{figure}

To reveal the full analytic properties of the S-matrix, we let the energy and momentum take complex values. With the relation in Eq.~\eqref{eq:nonrelE}, complex energy is a two-valued function of complex momenta with the positive or negative imaginary part. We call the energy plane associated with the positive (negative) imaginary part of the $i$th momentum channel as the top $[t]$ (bottom $[b]$) Riemann sheet. For the coupled two-channels, the four Riemann sheets are conveniently labeled using the intuitive notation in Ref.~\cite{PearceGibson} as $[s_1 s_2]$ where $s_i=\{t,b\}$ is the sheet of the $i$th channel. The scattering region, the energy region where we plot the cross-section, lies along the real axis on the upper half of the physical $[tt]$ sheet. The other Riemann sheets are collectively called unphysical sheets, and they are connected, together with the physical sheet, in a nontrivial way due to the presence of branch cut. Fig.~\ref{fig:riemann} shows the connection of the relevant regions of all the Riemann sheets in two perspectives. First, the upper-half of $[tt]$ ($[tb]$) is connected to lower-half of $[bt]$ ($[bb]$) between the interval $T_1$ to $T_2$. Fig.~\ref{fig:riemann}(a) shows that the upper-half of $[tt]$ is connected to the lower half of $[bb]$ above $T_2$. Second, Fig.~\ref{fig:riemann}(b) shows that the lower half of $[bt]$ is connected to the upper-half of $[tb]$ above $T_2$. 
Note that all the Riemann sheets are disconnected from each other below the lowest threshold $T_1$. 
Generally, poles near the scattering region, thus on the $[bb]$ and $[bt]$ sheets, may generate a prominent peak, but some poles on the $[tb]$ sheet can still give a noticeable effect, especially if close to the threshold energy.

The properties of the S-matrix govern the collision of particles with short-range interaction.
The most important property is causality, which means that no scattered wave will reach the detector before the incident wave reaches the scattering target. Causality imposes that the scattering amplitudes take the real-boundary values of some analytic functions of complex energy on the physical sheet~\cite{EdenAnalytic, Kampen1953}. It follows that there should be no pole on the physical sheet except possibly some bound state poles on the real axis below the lowest threshold. Next, we also expect that the total probability is conserved, which requires that the S-matrix $S$ be unitary, i.e., $SS^{\dagger}=1$. This property allows us to restrict the form of the full S-matrix and its elements in the corresponding interval of scattering energy. Finally, below the lowest threshold, the S-matrix should be real-valued or Hermitian~\cite{LandauQM}. The well-known consequence of hermiticity is the reflection principle~\cite{Taylor,Newton}, i.e., the energy poles come in conjugate pairs. By combining analyticity, unitarity, and hermiticity on the relevant energy region, we can construct a general parametrization for our S-matrix.

Using hermiticity and unitarity, we can perform analytic continuation~\cite{Newton1961,LeCouter1960} and obtain the following S-matrix elements: 
\begin{equation}
	\begin{split}
		S_{11}(p_1,p_2)&=\dfrac{D(-p_1,p_2)}{D(p_1,p_2)} \\
		S_{22}(p_1,p_2)&=\dfrac{D(p_1,-p_2)}{D(p_1,p_2)} \\
		S_{11}S_{22}-(S_{12})^2&=
		\dfrac{D(-p_1,-p_2)}{D(p_1,p_2)}
	\end{split}
	\label{eq:smatrix}
\end{equation}
where the subscripts refer to the channels. Here, the function $D(p_1,p_2)$ determines the pole positions of the S-matrix. We can calculate the scattering amplitudes using the relation $S_{ii'}=\delta_{ii'}+2iT_{ii'}$ where $\delta_{ii'}$ is the Kronecker delta and $i,i'\in\{1,2\}$.  If we can control the singularities using a specified form of $D(p_1,p_2)$, then we can generate a set of simulated amplitudes $T_{ii'}(p_1,p_2)$ with known pole configuration.

\subsection{Controlled poles and Riemann sheets}
\hspace{\parindent}
To accommodate a large pole-configuration space, we have to generate a set of simulated amplitudes with arbitrary number of independent poles. We look for $D_j(p_1,p_2)=0$ that can give exactly one pole $E^{(j)}_{\text{pole}}$ occupying one of the unphysical Riemann sheet and then form the product 
\begin{equation}
	D(p_1,p_2)=\prod_{j}D_j(p_1,p_2). 
\end{equation}
This prescription ensures that all $E^{(j)}_{\text{pole}}$s are produced independently of each other. Now, from the reflection principle, we know that if $E^{(j)}_{\text{pole}}$ is a pole, then the complex conjugate $E^{(j)*}_{\text{pole}}$ must also be a pole. To fix the Riemann sheet of $E^{(j)}_{\text{pole}}$ and accommodate its conjugate partner, we use
\begin{equation}
	\begin{split}
		D_j(p_1,p_2)=&\left[(p_1-i\beta_1^{(j)})^2-\alpha_1^{(j)2}\right] \\
		&+\lambda\left[(p_2-i\beta_2^{(j)})^2-\alpha_2^{(j)2}\right].
		\label{eq:Djp1p2}
	\end{split}
\end{equation}
The absolute values of the real parameters $\alpha_i^{(j)}, \beta_i^{(j)}$ are already determined by the real and imaginary parts of $E^{(j)}_{\text{pole}}$ through the relation in Eq.~\eqref{eq:nonrelE}. To specify the Riemann sheet of $E^{(j)}_{\text{pole}}$, we choose the signs of $\beta_1^{(j)}$ and $\beta_2^{(j)}$. This feature allows us to uphold the analyticity requirement by not choosing simultaneous positive $\beta_1^{(j)}$ and $\beta_2^{(j)}$. The importance of the extra parameter $\lambda$ is discussed in the following.

A quick inspection shows that, if Eq.~\eqref{eq:nonrelE} is imposed on $D_j(p_1,p_2)=0$ with $D_j(p_1,p_2)$ given by Eq.~\eqref{eq:Djp1p2}, we get four solutions in either $p_1$ or $p_2$. Two of these are the assigned pole $E^{(j)}_{\text{pole}}$ and its conjugate partner while the other two are known as the shadow and its conjugate partner \cite{Suzuki2009,Badalyan1982,Eden1964shadow,Frazer1964}. The position and Riemann sheet of the shadow pole is obtained by expressing $D_j(p_1,p_2)=0$ in terms of $p_1$ (or $p_2$) and factoring out $(p_1-i\beta_1^{(j)})^2-\alpha_1^{(j)2}$ (or $(p_2-i\beta_2^{(j)})^2-\alpha_2^{(j)2}$); these are
\begin{equation}
	\left[p_1-i\beta^{(j)}_1
	\left(\dfrac{\mu_1-\mu_2\lambda}
	{\mu_1+\mu_2\lambda}\right)
	\right]^2
	-
	\left[
	\alpha^{(j)2}_1+
	\dfrac{4\lambda\mu_1^2\beta^{(j)2}_2}
	{(\mu_1+\mu_2\lambda)^2}
	\right]
	=0
	\label{eq:shadowpoles1}
\end{equation}
and
\begin{equation}
	\left[p_2+i\beta^{(j)}_2
	\left(\dfrac{\mu_1-\mu_2\lambda}
	{\mu_1+\mu_2\lambda}\right)
	\right]^2-
	\left[
	\alpha_2^{(j)2}+\dfrac{4\lambda\mu_2^2\beta_1^{(j)2}}
	{(\mu_1+\mu_2\lambda)^2}
	\right]
	=0.
	\label{eq:shadowpoles2}
\end{equation}
The reversed sign of $\text{Im } p_2$ tells us that the Riemann sheet of the shadow is different to that of the assigned pole. This means that $E^{(j)}_{\text{pole}}$ can respect analyticity but the shadow may not. To avoid the possible violation of analyticity, we throw the shadow far from the relevant scattering region. In particular, we choose a sufficient negative $\lambda$ that can put the shadow (and its conjugate partner) on the real axis below the lowest threshold. By doing this, each $E^{(j)}_{\text{pole}}$ will have its own distant background pole. Thus, we can guarantee that each $D_j(p_1,p_2)$ will produce exactly one isolated pole on a specified Riemann sheet. In this way, the poles are produced independent of each other while respecting analyticity, giving us a general parametrization.

\subsection{Observable effects of shadow}
\hspace{\parindent}
We push the shadow pole solution of $D_j(p_1,p_2)=0$ far from the relevant scattering region to arrive at a general parametrization. However, there are situations where a shadow pole exists and may have observable consequences in the scattering region around the relevant threshold. Consider, for example; the familiar two-channel Breit-Wigner model \cite{Badalyan1982,Suzuki2009} with the pole-position condition taking the form
\begin{equation}
	D(p_1,p_2)=E-E_{\text{BW}}+i\gamma_1 p_1 + i\gamma_2 p_2 = 0
	\label{eq:twochannelBW}
\end{equation}
where $E_{BW}$ is the Breit-Wigner mass and $\gamma_i\geq 0$ is the coupling to the $i$th channel. For $\gamma_2<\gamma_1$, we can have a pole in $[bb]$ and a shadow in $[bt]$. If we slowly turn off $\gamma_2$, the pole and shadow will move towards a common energy point (the same real and imaginary part) but in different Riemann sheets. The cusp at $T_2$, due to $p_2\propto\sqrt{E-T_2}$, disappears on the scattering amplitude indicating that the situation is now reduced to a single-channel scattering. 

\begin{figure}
	\includegraphics[width=\columnwidth]{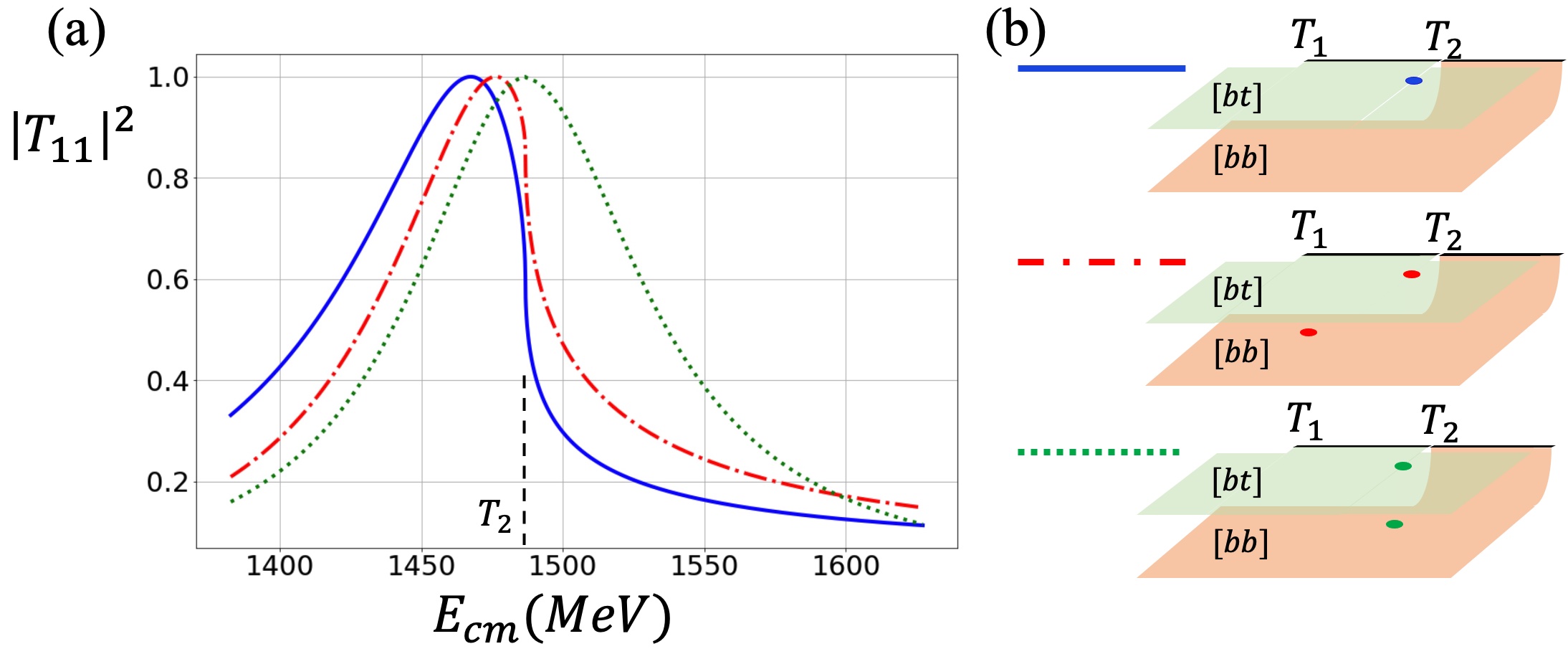}
	\caption{The partial cross-section (a) and the corresponding pole configuration (b). The pole in $[bt]$ sheet is the same for all cases with real part equal to $T_2$.}
	\label{fig:shadow}
\end{figure}

It turns out that putting another independent pole in our parametrization can mimic the effect of shadow. To demonstrate this effect, let us consider the extreme case where we fixed the position of an isolated pole on the $[bt]$ sheet such that its real part is equal to $T_2$. Fig.~\ref{fig:shadow} shows the resulting $|T_{11}|^2$ (solid blue line) where a peak structure below $T_2$ with a noticeable cusp (infinite slope) at the threshold is observed. The addition of an arbitrary $[bb]$ pole below $T_2$ push the peak structure slightly closer to $T_2$ (dashed red line) and slightly diminish the cusp structure. Finally, if the $[bb]$ pole is placed exactly at the same position as the fixed $[bt]$ pole, the peak occurs at the actual real part of the pole (which is at $T_2$), and the cusp disappeared (dotted green line). This behavior is an indication that the pole decouples to the second channel, similar to turning off $\gamma_2$ in the two-channel Breit-Wigner model. It follows that, we can turn off the channel coupling by using an arbitrary trajectory for two independent poles on different sheets. This behavior shows that the arbitrary pole we placed on a different Riemann sheet effectively functions as a shadow of the fixed pole. In our formulation, we can generate the pole and shadow independently. The decoupling effect of a pair of independent poles on different Riemann sheets can be proven in a general way as discussed in the Appendix.

We now have a parametrization that can produce as many poles as we want with an arbitrary way of controlling the strength of channel coupling. In the next section, we use our general S-matrix parametrization to produce a set of simulated amplitudes for the teaching dataset.

\section{Construction of deep neural network model}
\label{sec:dnnmodel}
\hspace{\parindent}
\subsection{Generation of training dataset}
The design of our DNN generally depends on the features of input data, which we take to be the real and imaginary parts of amplitude, and the number of possible classifications in the teaching dataset. In the final stage of analysis, we will use the experimental data to make DNN inferences. It is, therefore, essential to include the limited energy resolution in the design of DNN. To do this, we generate a set of amplitudes on randomly spaced energy points. Specifically, we divide the region of interest in the experimental data into some number of bins, say $B$ bins. One can think of each bin as the energy resolution of a particle detector. It is safe to assume that some probability distribution determines the energy of a particle entering a detector. We can use this distribution to pick one representative energy in each bin to calculate the real and imaginary parts of the amplitude. For our specific task, we divide the $\pi N$ center-of-mass energy range, from $\pi N$ to $K \Sigma$ thresholds, into $37$ bins since there are 37 data points in the GW-SAID data~\cite{SAIDpiN,SAIDpiNold}. Also, for simplicity, we use a uniform distribution to choose a point in each energy bin.

The number of possible pole-configuration classifications depends on how many poles we are willing to count within some specified Riemann sheet region. The number of configurations will also determine the number of output nodes assigned to our DNN model. Appealing once again to the GW-SAID data, there are two prominent structures - one is around the $\eta N$ threshold, and the other is between $K\Lambda$ and $K\Sigma$ thresholds. The number of poles that we have to count should not be less than two. In our study, it suffices to consider a maximum of four distributed poles in any combination of Riemann sheets. With this restriction, we identified 35 possible configurations. Each of these configurations corresponds to one output node in our DNN. The possible configurations and output-node assignment are shown in Table~\ref{tab:labelout}. 
 
 \begin{table}[ht!]
 	\centering
 	\caption{
 		Classification output-node label}
 	\begin{tabular}{c|l}
 		\hline
 		\textbf{Label}&
 		\textbf{S-matrix pole configuration} 
 		\\
 		\hline
 		0	&
 		no nearby pole
 		\\
 		\hline
 		1	&
 		1 pole in $[bt]$	
 		\\
 		2	&
 		2 poles in $[bt]$	
 		\\	
 		3	&
 		3 poles in $[bt]$ 
 		\\
 		4	&
 		4 poles in $[bt]$
 		\\
 		\hline
 		5	&
 		3 poles in $[bt]$ and 1 pole in $[bb]$ 
 		\\
 		6	&
 		2 poles in $[bt]$ and 1 pole in $[bb]$
 		\\
 		7	&
 		2 poles in $[bt]$ and 2 poles in $[bb]$
 		\\
 		8	&
 		1 pole in $[bt]$ and 2 poles in $[bb]$
 		\\
 		9	&
 		1 pole in $[bt]$ and 3 poles in $[bb]$
 		\\
 		10	&
 		1 pole in $[bt]$ and 1 pole in $[bb]$
 		\\
 		\hline
 		11	&
 		1 pole in $[bb]$
 		\\
 		12	&
 		2 poles in $[bb]$
 		\\
 		13	&
 		3 poles in $[bb]$
 		\\
 		14	&
 		4 poles in $[bb]$
 		\\
 		\hline
 		15	&
 		3 poles in $[bb]$ and 1 pole in $[tb]$ 
 		\\
 		16	&
 		2 poles in $[bb]$ and 1 pole in $[tb]$
 		\\
 		17	&
 		2 poles in $[bb]$ and 2 poles in $[tb]$
 		\\
 		18	&
 		1 pole in $[bb]$ and 2 poles in $[tb]$
 		\\
 		19	&
 		1 pole in $[bb]$ and 3 poles in $[tb]$
 		\\
 		20	&
 		1 pole in $[bb]$ and 1 pole in $[tb]$
 		\\
 		\hline
 		21	&
 		1 pole in $[tb]$
 		\\
 		22	&
 		2 poles in $[tb]$
 		\\
 		23	&
 		3 poles in $[tb]$
 		\\
 		24	&
 		4 poles in $[tb]$
 		\\
 		\hline
 		25	&
 		3 poles in $[tb]$ and 1 pole in $[bt]$ 
 		\\
 		26	&
 		2 poles in $[tb]$ and 1 pole in $[bt]$
 		\\
 		27	&
 		2 poles in $[tb]$ and 2 poles in $[bt]$
 		\\
 		28	&
 		1 pole in $[tb]$ and 2 poles in $[bt]$
 		\\
 		29	&
 		1 pole in $[tb]$ and 3 poles in $[bt]$
 		\\
 		30	&
 		1 pole in $[tb]$ and 1 pole in $[bt]$
 		\\
 		\hline
 		31	&
 		2 poles in $[bt]$, 1 pole in $[bb]$ and 1 pole in $[tb]$
 		\\
 		32	&
 		1 pole in $[bt]$, 2 poles in $[bb]$ and 1 pole in $[tb]$
 		\\
 		33	&
 		1 pole in $[bt]$, 1 pole in $[bb]$ and 2 poles in $[tb]$
 		\\
 		34	&
 		1 pole in $[bt]$, 1 pole in $[bb]$ and 1 pole in $[tb]$
 		\\
 		\hline		 				  	
 	\end{tabular} 
 	\label{tab:labelout}
 \end{table}
 
The enhancements in the GW-SAID $\pi N$ amplitude occur only on a limited portion of the scattering region. Thus, it is practical to limit the region where we produce and count our poles in each Riemann sheet. We define the counting region as:
\[
\begin{cases}
	T_2-50 \leq \text{ Re}E_{\text{pole}}\leq T_2+200
	& \text{all RS} \\
	-200\leq \text{Im }E_{\text{pole}}< 0 
	& \text{$[bt]\&[bb]$} \\
	0<\text{Im }E_{\text{pole}}\leq200 
	& \text{$[tb]$},
\end{cases}
\]
where the energies are in units of MeV.
Here, we do not have to count the conjugate poles on the upper half of $[bt]$ or $[bb]$ and the lower half of $[tb]$ since they do not correspond to different independent states. In addition to the counted poles, we also produce distant background poles far from the scattering region, but these are not counted. By randomly generating at-most four poles inside the counting region, we constructed $1.8\times 10^6$ labeled amplitudes divided uniformly into 35 configurations for the training set. Our task is to find a map between the input amplitude to the corresponding output pole configuration in the form of a DNN model. To monitor the performance of DNN, we also produce an independent $3.5\times 10^4$ labeled amplitudes for the testing set.

\begin{figure}[ht!]
	\includegraphics[width=\columnwidth]{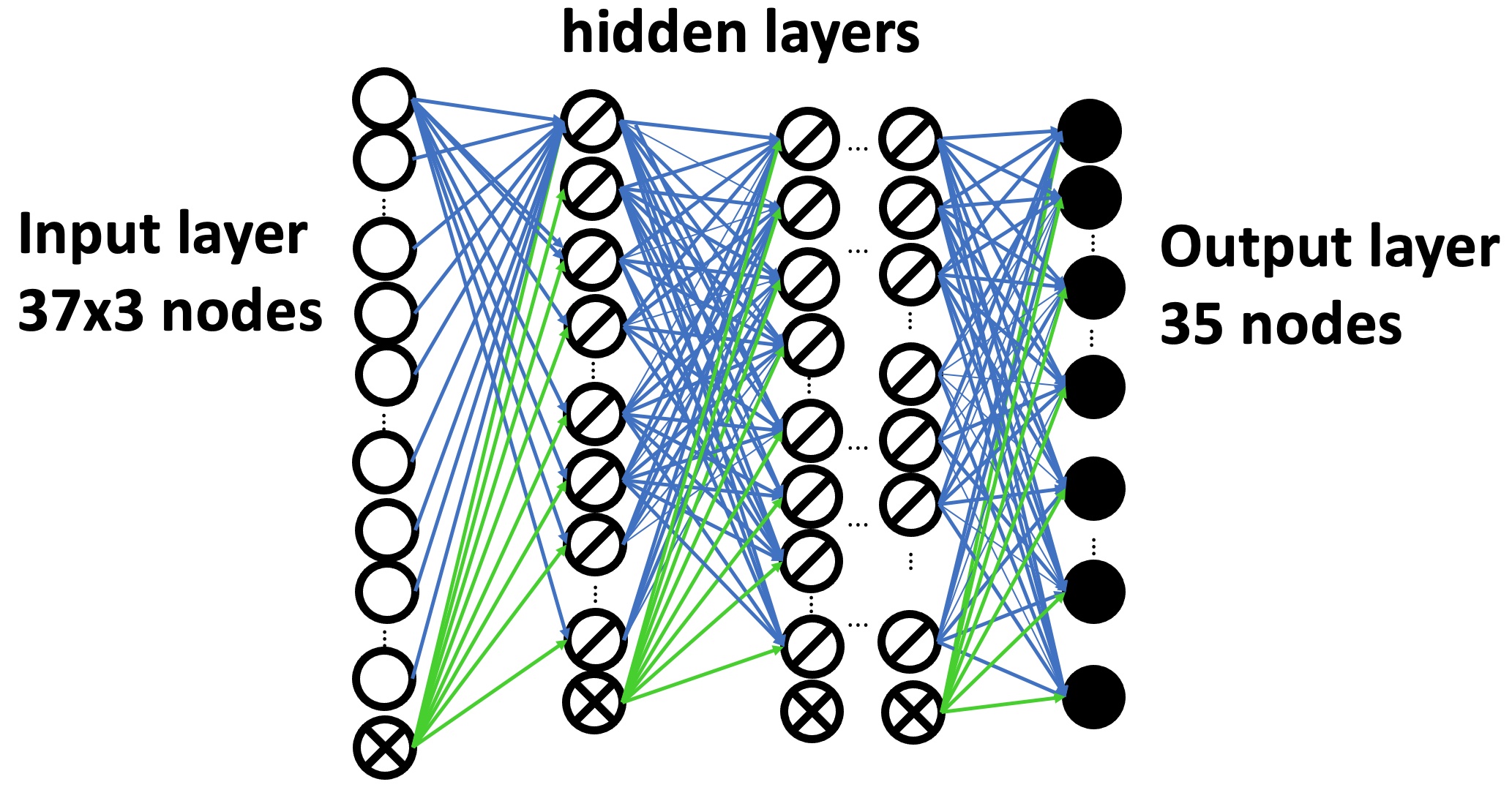}
	\caption{DNN architecture. The $\Circle$s are for input nodes, $\otimes$s are for bias nodes, $\oslash$s are for hidden layer nodes, and $\CIRCLE$s are for output nodes. The lines are the weights and biases.}
	\label{fig:dnnarchi}
\end{figure}
\begin{table}[h!]
	\centering
	\caption{
		Hidden layer architectures of DNN models considered in this study. The input and output layer nodes are identical for all models.}
	\begin{tabular}{c|c}
		\hline
		\textbf{DNN model}&
		\textbf{Hidden layer}	
		\\
		\textbf{label}&
		\textbf{architecture}	
		\\		
		\hline
		1					&
		[200-200]  								
		\\
		\hline
		2					&
		[200-200-200]  								
		\\
		\hline
		3					&
		[200-200-200-200]  								
		\\		
		\hline		
		4					&
		[100-100]  								
		\\		
		\hline	
		5					&
		[100-100-100]  								
		\\		
		\hline				
		6					&
		[100-100-100-100]  								
		\\		
		\hline											    
	\end{tabular}
	\label{tab:archi}
\end{table}

\subsection{DNN architecture}
Fig.~\ref{fig:dnnarchi} shows the basic architecture of our DNN model. The first 37 nodes in the input layer are for the random energy points chosen in each bin, and the last two 37 nodes are for real and imaginary parts of the amplitude. For the output layer, we use 35 nodes such that each one corresponds to the pole configuration listed in Table~\ref{tab:labelout}. There are no general guidelines into how many hidden layers and nodes in each layer we should assign to obtain an optimal result. It is, therefore, useful to test some architectures. In this study, we experiment with six different architectures with hidden layer designs shown in Table~\ref{tab:archi}.

Except for the input layer nodes and the biases, all the other nodes are equipped with an activation function. For the hidden layer nodes, we use the rectified linear unit (ReLU) such that for the $n$th node
\begin{equation}
	\mbox{ReLU}\left(z_n^{(N+1)}\right)
	=\mbox{max}\left(0,z_n^{(N+1)}\right)
	\label{eq:relu}
\end{equation}  
where $z_n^{(N+1)}$ is the linear combination of all the $N$th layer node values multiplied by the appropriate weights and shifted by the bias of the same layer. For the output layer, it is optimal to use the softmax given by
\begin{equation}
	\mbox{softmax}\left(z_n^{(L+1)}\right)=\dfrac{\exp\left(z_n^{(L+1)}\right)}{\sum_{m}^{N_{L+1}}\exp\left(z_m^{(L+1)}\right)}
	\label{eq:softmax}
\end{equation}      
where $L$ is the last of the hidden layers. We also use the softmax cross entropy as our cost-function. The construction of the DNN models and the execution of training loop are all done in Chainer~\cite{Chainer2015,Chainer2017,Chainer2019,ChainerGithub}.

In a typical training loop, we feed the teaching dataset using some mini-batch procedure to add stochasticity in estimating the cost-function. Then, we execute the optimization of weights and biases using some variant of stochastic gradient descent~\cite{Adam}. We perform these procedures on all our models and found that none is learning the classification problem. In particular, the training and testing accuracies remain at around $2.86\%$ no matter how many training epochs we use. The observed accuracy is, in fact, the accuracy that we will get if we make a random guess out of 35 possibilities. We tried different optimizers and different combinations of hyperparameters shown in Table~\ref{tab:noncurr} but still obtained a non-improving performance. 

\begin{table}[b!]
	\centering
	\caption{List of optimizers and hyperparameters used in the non-curriculum traning. For further descriptions see Ref.~\cite{ChainerGithub}}
	\begin{tabular}{c|l}
		\hline
		{Optimizers}&
		{Adam, AdaDelta, AdaGrad} \\
		{}& 
		{AMSGrad, AdaBound, AMSBound} \\
		\hline
		{Mini-batch sizes}&
		{32, 64, 512, 1024, 1536} \\
		{}& 
		{2048, 2560, 4096} \\
		\hline
		{Weight Initializers}&
		{Normal, HeNormal, Uniform} \\
		{}& 
		{HeUniform, Orthogonal} \\
		\hline			
		{Others}&
		{with Dropout, without Dropout} \\
		\hline					
	\end{tabular} 
	\label{tab:noncurr}
\end{table}	

The difficulty in learning the classification problem is due to the noise introduced in the generation of the training dataset. This noise includes the random choice of energy points in evaluating amplitudes and the randomly added unitary background poles. It is often advisable to use a dataset with less noise to improve the training performance~\cite{Aggarwal2018}, but this defeats the purpose of simulating the energy resolution in the experimental data. The inclusion of energy resolution in the design of DNN inevitably results in a noisy dataset. Thus, we resort to a different approach to initiate the learning process without tampering with our teaching dataset. In the following, we introduce the idea of the curriculum method and how it can start the learning of a complicated classification problem.

\subsection{Curriculum method}
\hspace{\parindent}
The curriculum method was first developed in Ref~\cite{ELMAN199371} for simple cases. Depending on the difficulties of classification problems, a more rigorous treatment requires a well-organized training dataset~\cite{CurriculumLearning0, CurriculumLearning1,CurriculumLearning2}. For our purpose, it is sufficient to adopt a more heuristic approach where we subjectively identify the simplest dataset and introduce new classification until we present all the training sets. With the 35 pole-configuration classifications, four of which corresponds to at-most-one-pole configuration (labels 0, 1, 11, and 21 of Table~\ref{tab:labelout}). We treat these four classes as the simplest examples and call them curriculum 1. Then we add one of the two-pole classifications and call the new set curriculum 2. We do this until we have included all the 35 classifications in the final curriculum 32. Table~\ref{tab:curriculum} shows the incremental progression of the curriculum training. Note that by the end of curriculum 7, we have introduced all the two-pole configurations. Similarly, all the three-pole configurations are presented at curriculum 17 and all four-pole configurations at curriculum 32. The code used to build each curriculum is in the public repository~\cite{MyGithub2}.
 \begin{table}[ht!]
	\centering
	\caption{
		Scheme that we used in the curriculum learning stage. 
		One new classification label is added in the subsequent curriculum. The label descriptions are given in
		Table~\ref{tab:labelout}
	}
	\begin{tabular}{c|l|l}
		\hline 
		\textbf{Curriculum}&
		\textbf{Dataset addition} &
		\textbf{Configuration}\\
		\textbf{Label}&
		\textbf{Scheme}&
		\textbf{presented}
		\\
		\hline
		1	&
		Labels 0,1,11,21 &
		at-most 1 pole
		\\
		\hline
		2	&
		Curriculum 1 + Label 2	&
		\\
		3	&
		Curriculum 2 + Label 12 &
		\\	
		4	&
		Curriculum 3 + Label 22 &
		\\
		5	&
		Curriculum 4 + Label 10 &
		\\
		6	&
		Curriculum 5 + Label 20 &
		\\
		7	&
		Curriculum 6 + Label 30 &
		at-most 2 poles
		\\
		\hline
		8	&
		Curriculum 7 + Label 3 &
		\\
		9	&
		Curriculum 8 + Label 13 &
		\\
		10	&
		Curriculum 9 + Label 23 &
		\\
		11	&
		Curriculum 10 + Label 6 &
		\\
		12	&
		Curriculum 11 + Label 8 &
		\\
		13	&
		Curriculum 12 + Label 16 &
		\\
		14	&
		Curriculum 13 + Label 18 &
		\\
		15	&
		Curriculum 14 + Label 26 &
		\\
		16	&
		Curriculum 15 + Label 28 &
		\\
		17	&
		Curriculum 16 + Label 34 &
		at-most 3 poles
		\\
		\hline
		18	&
		Curriculum 17 + Label 4 &
		\\
		19	&
		Curriculum 18 + Label 14 &
		\\
		20	&
		Curriculum 19 + Label 24 &
		\\
		21	&
		Curriculum 20 + Label 5 &
		\\
		22	&
		Curriculum 21 + Label 7 &	
		\\
		23	&
		Curriculum 22 + Label 9 &
		\\
		24	&
		Curriculum 23 + Label 15 &
		\\
		25	&
		Curriculum 24 + Label 17 &
		\\
		26	&
		Curriculum 25 + Label 19 &
		\\
		27	&
		Curriculum 26 + Label 25 &
		\\
		28	&
		Curriculum 27 + Label 27 &
		\\
		29	&
		Curriculum 28 + Label 29 &				
		\\
		30	&
		Curriculum 29 + Label 31 &
		\\
		31	&
		Curriculum 30 + Label 32 &
		\\						
		32	&
		Curriculum 31 + Label 33 &
		at-most 4 poles
		\\						
		\hline		 				  	
	\end{tabular} 
	\label{tab:curriculum}
\end{table}
\begin{figure*}
	\includegraphics[width=0.9\linewidth]{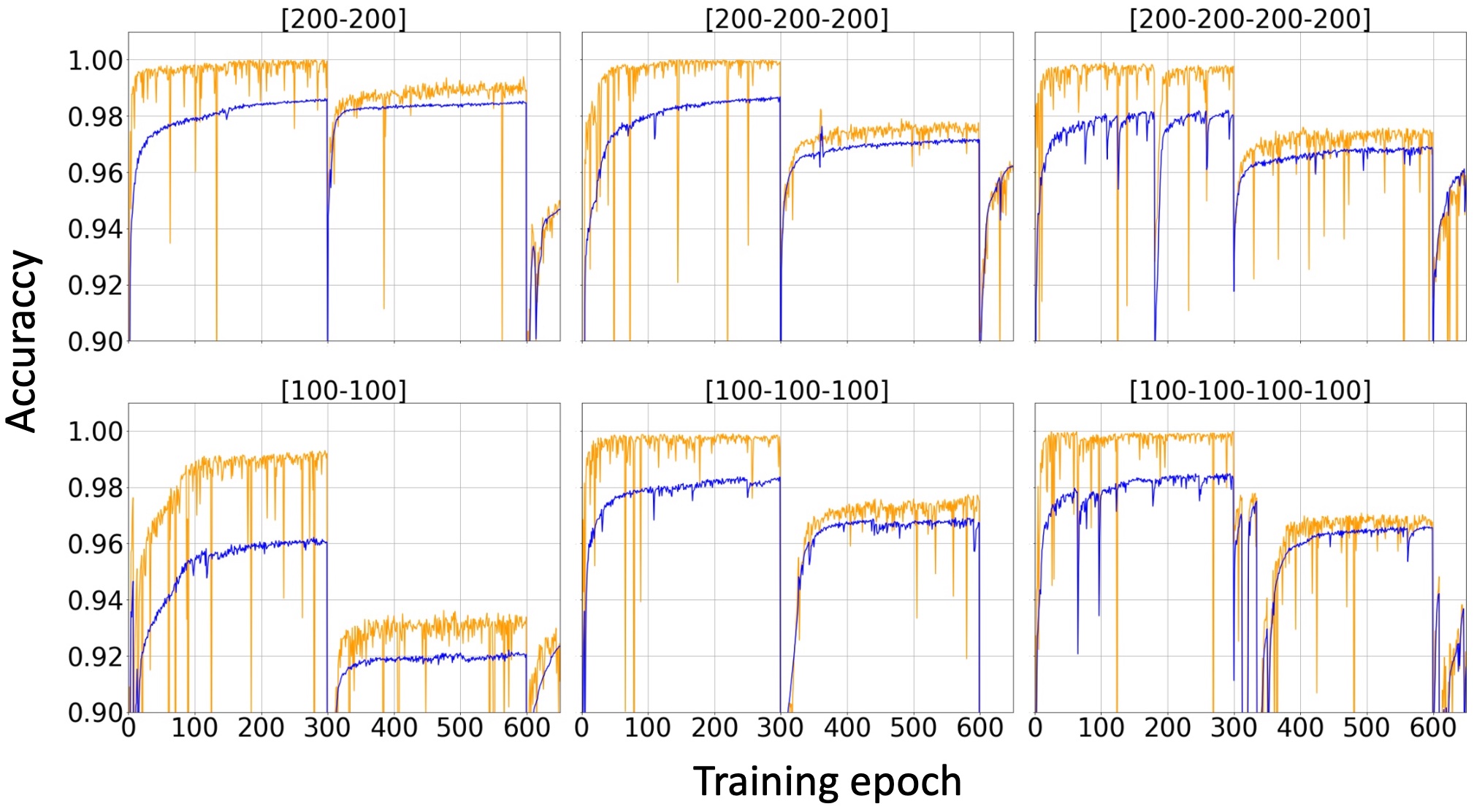}
	\caption{Performance of the DNN models for the first three curricula. The blue lines are for the training performance while the orange lines are for the testing performance.}
	\label{fig:multidnn}
\end{figure*}

We trained all the six models using curriculum 1 for the first 300 epochs, curriculum 2 for the subsequent 300 epochs, and then curriculum 3 until epoch 650. The goal here is to find which architecture performs best and then devote the rest of the training to the chosen DNN. Fig.~\ref{fig:multidnn} shows the performance of the six DNNs. All models give decent accuracies at the start of curriculum 1, indicating that the chosen simple classification is indeed learnable. There is a noticeable accuracy drop at the start of curriculum 2 and curriculum 3 due to the introduction of a new classification set. Nevertheless, we can perform a few more training epochs within the curriculum to improve accuracy. The essential point here is that all our models start to learn the classification problem using the curriculum approach. Of all the six models, the architecture with a hidden layer [200-200-200] shows a promising performance, especially at the onset of curriculum 3. We, therefore, devote the rest of our computing resources to DNN model 2. 

Note that it is not practical to perform more epoch per curriculum since the accuracy will definitely drop at the start of a new curriculum. It will suffice to have some decent accuracy for each curriculum and continue the training after introducing all the classifications. Hence, we restart the training of the chosen DNN and, to accelerate the process, we only use 100 epochs per curriculum which are then continued after all the pole configurations are introduced. We also vary the mini-batch size from 512 in the early part of the curriculum epoch, so we can take advantage of large stochasticity to 4,608 in the later part to stabilize the accuracy. At the end of epoch 3,200 (end of curriculum 32), we now have a DNN model that can detect up to four poles in any Riemann sheet with training and testing accuracies of $63.5\%$ and $68.3\%$, respectively. We continue until epoch 31,050 using the typical training loop and obtained a final training and testing accuracies of $76.5\%$ and $80.4\%$, respectively~\cite{SombilloPRL2021}. 

Recall that in the construction and training of our DNN model, only the energy resolution is included and not the fluctuation of the amplitudes, which correspond to statistical errors in experimental data. The goal of using a generic S-matrix is to teach the DNN to recognize only those amplitudes that satisfy the unitarity, analyticity, and hermiticity requirements. This restriction is justified because we expect the actual experimental data to conform to the mentioned general properties. Inserting a random error or offset in the training amplitudes will violate such requirements. In the inference stage, the trained DNN will reinterpret the input experimental amplitudes as if they satisfy the expected properties even if there are some offsets. This feature is the advantage of having a model with many parameters (weights and biases) where the trained DNN automatically discards the irrelevant peculiarities of the input data. We further describe the inference stage in the next section.    

\section{Application to $\pi N$ scattering}
\label{sec:apply}
\hspace{\parindent}
We can now use our trained DNN to make inferences on the experimental data. 
Due to the presence of error bars of the data, we can expect that there are multiple possible interpretations associated with the amplitude of interest. The result of DNN inferences must reflect the uncertainty due to the error bars in the experimental data. We can accomplish this by combining points in each error bar to produce several amplitudes. Here, we can further assume that some probability distribution weights the points in each error bar. Except for the assumed probability distribution, the generated amplitudes for the DNN inference comes directly from the experimental data without imposing anything.

\begin{figure}[t!]
	\includegraphics[width=\columnwidth]{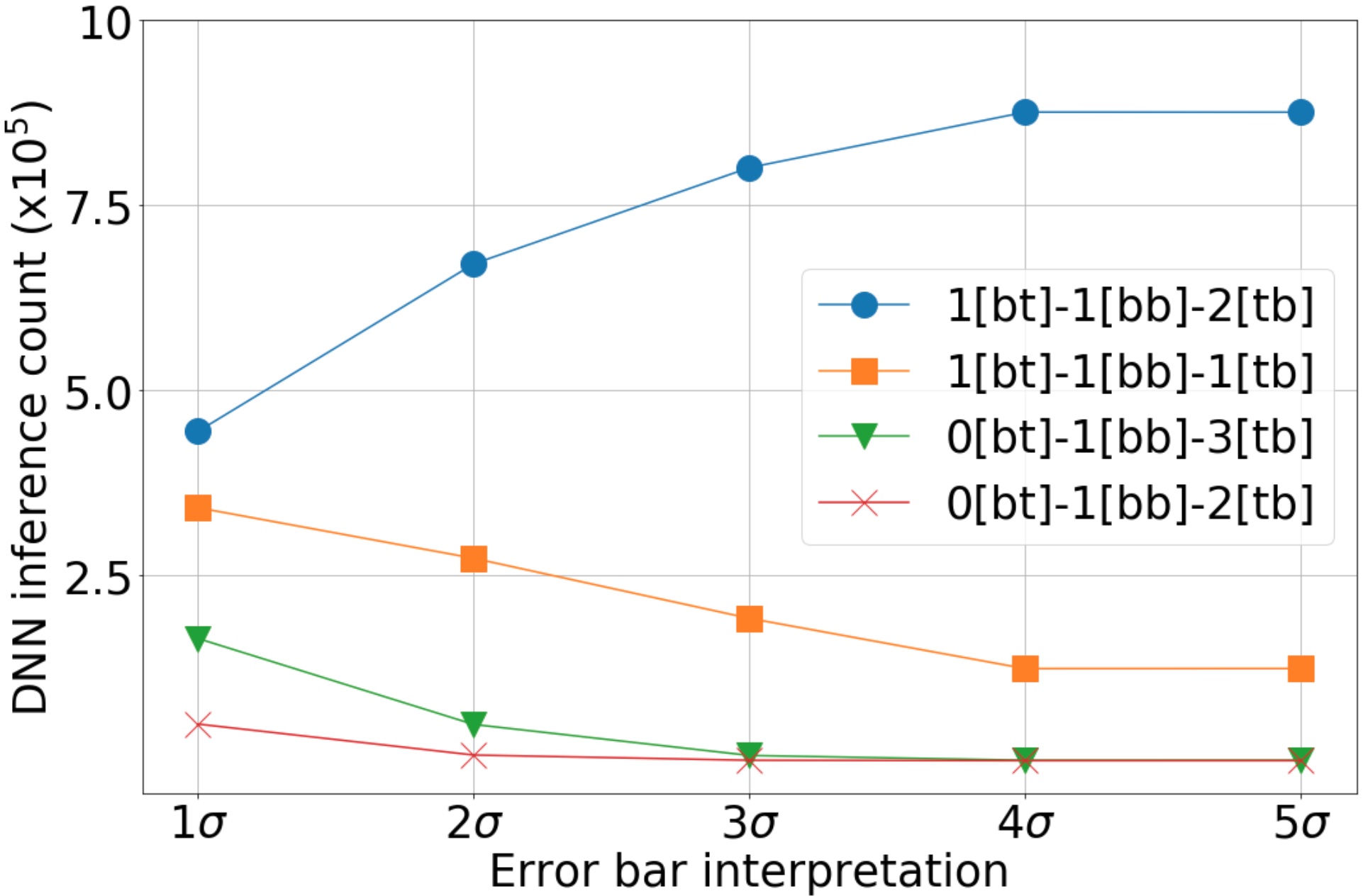}
	\caption{DNN inference on amplitudes generated from experimental data with points drawn from Gaussian distribution in each error bar. The legend shows the identified pole configurations by the DNN.}
	\label{fig:dnninfgauss}
\end{figure}

\subsection{Interpreting the error bar}
\hspace{\parindent}
We usually interpret each error bar to be one standard deviation $\sigma$ of the Gaussian distribution around the central value. Therefore, we can generate $10^6$ amplitudes directly from the experimental data by drawing points in each error bar using the Gaussian distribution. Note that the way we interpret the error bar corresponds to the confidence level that the points used to generate the amplitudes are within the error bar of the data. For example,  increasing the $\sigma$ to $2\sigma$ means that we increase the confidence level from $68\%$ to $95\%$, and so on. In Fig.~\ref{fig:dnninfgauss} we use different interpretations of the error bar in the generation of $10^6$ amplitudes and then count the number of DNN inference output. The result shows that out of 35 possibilities, only four-pole configurations are identified by the DNN. These detected configurations are shown in the legend of Fig. 6. Notice that as we increase the confidence level of the error bar, the most likely configuration emerges. Specifically, we find that the structures in the elastic $\pi N$ amplitude are caused by one pole in $[bt]$, one pole in $[bb]$, and two poles in $[tb]$. The other three configurations are suppressed as the confidence level is increased, implying that these three configurations correspond to amplitudes produced by points outside the error bars. 

\begin{table}[b!]
	\centering
	\caption{Result of the DNN inferences on the GW-SAID $\pi N$ amplitude. The points used to generate the $10^6$ experimental amplitudes are drawn in each error bar using a uniform distribution. We inform the reader that this table is different to Table II of~\cite{SombilloPRL2021}.}
	\begin{tabular}{c|c|c|c}
		\hline
		\textbf{Percentage}&
		\textbf{bt}&
		\textbf{bb}&
		\textbf{tb} \\
		\hline
		60.3\%&
		1&
		1&
		2\\
		30.9\%&
		1&
		1&
		1\\		
		7.5\%&
		0&
		1&
		3\\	
		1.3\%&	
		0&
		1&
		2\\
		\hline		
	\end{tabular} 
	\label{tab:uniform}
\end{table}	

One advantage of our approach is that we can go beyond the typical Gaussian distribution of points in each error bar and use other distributions. In particular, all the points in each error bar may be equally important, and a uniform distribution might be appropriate. In this way, all points used to generate the amplitudes are guaranteed to remain within the error bar. The DNN inference result, with the uniformly distributed points in each error bar, is shown in Table~\ref{tab:uniform}. It is interesting to note that changing the weights at which the points are combined to produce the amplitudes does not affect the DNN inference very much. That is, we still get the same conclusion that the $\pi N$ amplitude is best described by one pole in each adjacent Riemann sheet, one at most two poles in the distant sheet. Our approach contrasts with the conventional model-fitting scheme, where only a selected region of each error bar is used to describe the data. Thus, our deep learning approach is statistically robust compared to the conventional model-fitting scheme.

\subsection{Discussion of results}
\hspace{\parindent}
Note that we made no assumptions on the poles detected by the DNN.
At this point, one can now use a model to interpret the origin of the detected poles. Here, we give some general statements based on the expected effects of Riemann sheet poles on the scattering amplitude. First, the most prominent structure in Fig.~\ref{fig:said_data} is the enhancement between the $K\Lambda$ and $K\Sigma$ thresholds. For the two-channel analysis, such enhancement can only be produced by at least one $[bb]$ pole above the second threshold. Meaning, we can associate this enhancement to the $[bb]$ pole that our trained DNN consistently identifies.

The next noticeable structure is around the $\eta N$ threshold. As an intuitive interpretation, one may associate the detected $[bt]$ pole to this near-$\eta N$ structure. However, a near-threshold $[bt]$ pole is expected to give rise to an amplitude peak below the threshold with an imaginary part close to unity, i.e., reaching the unitarity limit. This description is not the case with the near-$\eta N$ enhancement. Also, notice that the DNN sometimes identifies the $[bt]$ pole as a $[tb]$ pole (compare the first and the third rows of Table~\ref{tab:uniform}). The comparison suggests that the $[bt]$ pole is near the $[bt]$-$[tb]$ interface which is located above the $\eta N$ threshold (see Fig.\ref{fig:riemann}(b)). The detected $[bt]$ pole may be associated with the enhancement between the $K\Lambda$ and $K\Sigma$ thresholds in the form of a shadow pole. Thus, we can only attribute the $\eta N$ threshold enhancement to a non-$[bt]$ sheet pole.

The enhancement around $\eta N$ threshold could be due to the threshold cusp effect. However, to make the structure prominent, some nearby poles must be associated with it. Notice that our trained DNN consistently detected $[tb]$ poles in all of its inferences. With the $[bb]$ pole already allocated to the other enhancement and with $[bt]$ pole ruled out, we are only left with at most two poles in the $[tb]$ sheet to account for the $\eta N$ threshold enhancement. The limitation imposed on our counting region guarantees that at least one of the detected $[tb]$ poles is close to the $\eta N$ threshold. The peak structure slightly above the $\eta N$ threshold suggests that one of the $[tb]$ poles is close to the $[bb]$-$[tb]$ interface.

We can only make further statements about the origin of the detected poles by appealing to some dynamical model. Nevertheless, we can give some general speculations on the nature of the detected poles. It might be the case that in the zero-coupling limit of $\pi N$ and $\eta N$ channels, the detected $[bb]$ and the $[bt]$ poles move towards a common position resulting in a typical Breit-Wigner resonance of the lower channel. Such is a typical feature of a pole-shadow pair. On the other hand, in the zero coupling limit, one of the $[tb]$ poles go to the $[bb]$ sheet and approaches the same position as the other $[tb]$ pole. In this way, the pole decouples the lower channel and results in either a near-threshold resonance or just a virtual state of the higher channel. Note that a dynamical model is needed to verify the above speculations.

\section{Conclusion and Outlook}
\label{sec:conc}
\hspace{\parindent}
We have demonstrated how to design a DNN that can extract the pole configuration of a given scattering amplitude. Using a generic S-matrix, we can produce a sizable teaching dataset to train our model to detect the number of poles in each Riemann sheet. Pole generation using Eqs.\eqref{eq:smatrix}-\eqref{eq:Djp1p2} gives us the advantage of performing deep learning analysis on any elements of the S-matrix. This prescription allows us to use the method on any scattering processes such as $2\rightarrow2$ or $1\leftrightarrow2$. Our approach can be extended to higher channels by putting an extra $\lambda\left[\left(p-i\beta\right)^2-\alpha^2\right]$ term. Here, the new parameter $\lambda$ can be adjusted to control the new shadow produced by the new channel. Additionally, the uniformization method introduced in Refs.~\cite{Yamada2020,Yamada2021} also provides an alternative way to control the poles and generate the training dataset. One possible advantage of uniformization is that it can be extended to relativistic scattering; this will be considered elsewhere.

Also, we have shown the effectiveness of the curriculum method in initiating the learning process with a noisy dataset.
The curriculum method allows us to accommodate the limited energy resolution in the design of DNN. We also provided a method of utilizing the error bars in the experimental data in a statistically robust way. By generating several experimental amplitudes, we can make a reasonable interpretation of the data based on the most consistent inference of the DNN.

Finally, our proposed S-matrix treats the experimental results in a model-independent way. The poles produced for the training dataset are independent of each other and are not constrained by any a priori trajectory. 
The above implies that the trained DNN makes no assumptions in detecting the poles associated with the experimental data.
It is now up to some dynamical model to interpret which set of poles are supposed to be paired as pole-shadow partners or which one is independent of the other. 

\section*{Acknowledgment}
This study was supported in part by MEXT as “Program for Promoting Researches on the Supercomputer Fugaku” (Simulation for basic science: from fundamental laws of particles to creation of nuclei).
DLBS is supported in part by the DOST-SEI ASTHRDP postdoctoral research fellowship. 
YI is partly supported by JSPS KAKENHI Nos. JP17K14287 (B) and 21K03555 (C).
AH is supported in part by JSPS KAKENHI No. JP17K05441 (C) and Grants-in-Aid for Scientific Research on Innovative Areas, No. 18H05407 and 19H05104.

\section*{Appendix}
\label{sec:append}
\hspace{\parindent}
We show that an irrelevant independent pole can remove the cusp at the higher threshold of a coupled channel scattering. Note that the origin of a threshold cusp is purely kinematical. This will occur at $E=T_2$ when the amplitude has an explicit dependence on $p_2$, where $p_2\propto\sqrt{E-T_2}$. If we can manage to remove the $p_2$ (like turning off $\gamma_2$ in two-channel Breit-Wigner) or turn $p_2$ into some analytic function of $E$ in the amplitude, then the cusp singularity should disappear. 

Let the relevant pole $E_{\text{pole}}$ in, say, $[bb]$ sheet (i.e. $\text{Re}E_{\text{pole}}>T_2$) satisfies the condition:
\begin{equation*}
	F(E_{\text{pole}}) + i g_1(E_{\text{pole}}) p_1 + i g_2(E_{\text{pole}}) p_2 = 0
	\label{eq:polecondition}
\end{equation*} 
where $F, g_1, g_2$ are analytic functions of $E$. We can always find $F, g_1, g_2$ such that the nearest singularity to the scattering region is an isolated simple pole. Since $E_{\text{pole}}$ is in $[bb]$, then we can express the momentum poles $(p_1, p_2)$ as
\begin{equation*}
	\begin{split}
		p_1 &= -i\beta_1\pm\alpha_1 \\
		p_2 &=-i\beta_2\pm\alpha_2
	\end{split}
\end{equation*}
where we set $\alpha_1, \alpha_2, \beta_1$ and $\beta_2$ to be positive. It is understood that $p_1$ and $p_2$ are related to the $E_{\text{pole}}$ via the energy constraint in Eq.~\eqref{eq:nonrelE}. It follows that, the modified equation 
\begin{equation*}
	F(E_{\text{pole}}) + i g_1(E_{\text{pole}}) p_1 - i g_2(E_{\text{pole}}) p_2 = 0.
	\label{eq:polecondition2}
\end{equation*} 
is satisfied by the same $E=E_{\text{pole}}$ but in the $[bt]$ sheet, i.e. the momentum poles are 
\begin{equation*}
	\begin{split}
		p_1 &= -i\beta_1\pm\alpha_1 \\
		p_2 &=i\beta_2\mp\alpha_2.
	\end{split}
\end{equation*}

Now, consider the S-matrix element $S_{11}(p_1, p_2)$ given by
\begin{equation*}
	\begin{split}
		S_{11}(p_1,p_2)=&
		\left(\dfrac{F - i g_1 p_1 + i g_2 p_2}
		{F + i g_1 p_1 + i g_2 p_2}\right) \\
		&\times
		\left(\dfrac{F - i\bar{g_1} p_1 - i\bar{g_2} p_2}
		{F + i\bar{g_1} p_1 - i\bar{g_2} p_2}\right).
	\end{split}
\end{equation*}
The above form is consistent with $S_{11}(p_1,p_2)=D(-p_1,p_2)/D(p_1,p_2)$ in Eq.~\eqref{eq:smatrix}. From the previous discussion, we know that the first parenthetical factor will generate a relevant pole $E_{pole}$ in $[bb]$ sheet. 

Deform $\bar{g_1}$ and $\bar{g_2}$ arbitrarily such that $\bar{g_1}\rightarrow g_1$ and $\bar{g_2}\rightarrow g_2$, respectively. The second factor will produce an irrelevant pole at $E=E_{\text{pole}}$ in $[bt]$ sheet. Using this limiting procedure, the S-matrix element becomes
\begin{equation*}
	S_{11}(p_1,p_2)=
	\dfrac{(F - i g_1 p_1)^2 + (g_2 p_2)^2}
	{(F + ig_1 p_1)^2 + (g_2 p_2)^2}.
\end{equation*}
The explicit dependence of $S_{11}(p_1,p_2)$ on $p_2$ is now replaced with $p_2^2$. We still have a peak at around $E=\text{Re}E_{\text{pole}}$ due to the relevant pole since the denominator did not cancel out as we perform the limiting procedure. However, the amplitude will no longer have a branch cut or threshold cusp at $E=T_2$ due to the absence of explicit $p_2$ dependence. Furthermore, from Eq.~\eqref{eq:smatrix}, the limit
$\bar{g_1}\rightarrow g_1$ and $\bar{g_2}\rightarrow g_2$, will give us $S_{22}\rightarrow1$ and $S_{12}^2\rightarrow0$. This means that putting an arbitrary irrelevant pole in the same position as the main pole, but on a different Riemann sheet, effectively decouples the two channels.

\bibliography{mybib}

\end{document}